\documentclass[npg]{copernicus}

\begin{document}

\title{The Onsager--Machlup functional for data assimilation}

\Author{Nozomi}{Sugiura}

\affil{Research and Development Center for Global Change, JAMSTEC, Yokosuka, Japan}

\runningtitle{OM functional for data assimilation}

\runningauthor{N. Sugiura}

\correspondence{Nozomi Sugiura (nsugiura@jamstec.go.jp)}

\received{26 July 2017}
\pubdiscuss{27 July 2017}
\revised{26 October 2017}
\accepted{26 October 2017}
\published{}

\hack{\allowdisplaybreaks}

\firstpage{1}

\texlicencestatement{This work is distributed under \hack{\newline} the Creative Commons Attribution 4.0 License.}
\maketitle

\nolinenumbers

\begin{abstract}
  When taking the model error into account in data assimilation,
  one needs to evaluate the prior distribution
  represented by the Onsager--Machlup functional.
  Through numerical experiments, this study clarifies how
 the prior distribution should
  be incorporated into cost functions for discrete-time estimation problems.
  Consistent with previous theoretical studies, the divergence of the drift term
  is essential in weak-constraint 4D-Var (w4D-Var),
but it is not necessary in Markov chain Monte Carlo with the Euler scheme.
  Although the former property may cause difficulties when implementing
  w4D-Var in large systems,
  this paper proposes a new technique
  for estimating the divergence term and its derivative.
\end{abstract}

\introduction  \label{Intro}

In traditional weak-constraint 4D-Var settings
\citep[e.g.][]{zupanski1997general,QJ:QJ200613262101}, a quadratic cost
function is defined as the negative logarithm of the probability for each
sample path, which is suitable for path sampling
\citep[e.g.][]{zinn2002quantum}. The optimisation problem is naively
described as finding the most probable path by minimising the quadratic cost
function. However, the term ``the most probable path'' does not make sense in
this context, because the paths are not countable. One should note that the
concern is not about ranking the individual path probabilities, but about
seeking the route with the densest path population. To define the
optimisation problem properly, one should introduce a macroscopic variable
$\phi=\phi(t)$ that represents a smooth curve, and introduce a measure that
accounts for how densely the paths are populated in the
$\epsilon$-neighbourhood centred at $\phi$, which can be termed ``the tube''.
Then the problem is defined as finding the most probable tube $\phi$, which
represents the maximum a posteriori (MAP) estimate of the path distribution.
Mathematicians pioneering the theory of stochastic differential equations
(SDEs) \citep[e.g.][]{ikeda2014stochastic,zeitouni1989onsager} have been
aware of this subtle point since the 1980s, and established the proper form
of the cost function as the Onsager--Machlup (OM) functional
\citep{onsager1953fluctuations} for the tube.

The aim of this work is to organise existing knowledge about the OM
functional into a form that can be used to represent model errors in data
assimilation, i.e. numerical evaluation of non-linear smoothing problems.

Throughout this article, we consider non-linear smoothing problems of the
form
\begin{align}
  \mathrm{d}x_t& = f(x_t) \mathrm{d}t + \sigma \mathrm{d}w_t, \label{SDE1}\\
  x_0&\sim \mathcal{N}(x_\mathrm{b},\sigma_\mathrm{b}^2 I), \label{SDE1bg}\\
 (\forall m \in M)\quad y_m|x_m &\sim \mathcal{N}(x_m,\sigma_\mathrm{o}^2 I),
\label{SDE1obs}
\end{align}
where $t$ is time, $x$ is a $D$-dimensional stochastic process, $w$
is a $D$-dimensional Wiener process, $x_\mathrm{b} \in \mathbb{R}^D$ is the
background value of the initial condition, $\sigma_\mathrm{b}>0$ is the
standard deviation of the background value, $y_m \in \mathbb{R}^D$ is
observational data at time $t_m$, $x_m=x_{t_m}, ~ t_m= m\delta_t$, $M$ is the
set of observation times, $\sigma_\mathrm{o}>0$ is the standard deviation of
the observational data, and $\sigma>0$ is the noise intensity. Note that
there is no need to distinguish the Ito integral from the Stratonovich
integral with regard to the discretisation of the SDE, because the noise
intensity is a constant.

Before moving on to its applications, here we review the concept of the OM
functional. To make presentation simple, we assume that $D=1$ and $\sigma
=1$, and concentrate on the formulation of the prior distribution in the
subsequent two Sects.~\ref{OM_path} and \ref{OM_MAP}. \hack{\newpage}
\subsection{OM functional for path sampling}\label{OM_path}

The model Eq.~(\ref{SDE1}) is discretised with the Euler scheme (with the
drift term at the previous time) as
\begin{align}
  x_n&=x_{n-1}+f(x_{n-1})\delta_t + \xi_{n-1},\quad
  n=1,2,\cdots,N,\label{disc_x}
\end{align}
where $\delta_t$ is the time increment, and each $\xi_{n-1}$ obeys
$\mathcal{N}(0,\delta_t)$. Equation~(\ref{disc_x}) can be considered a
non-linear mapping $F_1: {\xi} \mapsto x$ from the noise vector
${\xi}=({\xi}_0,{\xi}_1,\cdots,{\xi}_{N-1})^T$ to the state vector
$x=(x_1,x_2,\cdots,x_{N})^T$. The inverse of the mapping is linearised as
\begin{align}
 & \begin{bmatrix}
    \delta {\xi}_0\\
    \delta {\xi}_1\\
    \vdots\\
    \delta {\xi}_{N-1}
  \end{bmatrix}
= \nonumber\\ &
  \begin{bmatrix}
    1                        & 0&\cdots &                             0 & 0 \\
    -1-\delta_t f'(x_1)& 1&       &                             0 & 0 \\
    \vdots                   &  &       &                               &  \vdots \\
    0                        & 0&\cdots & -1-\delta_t f'(x_{N-1}) & 1
  \end{bmatrix}\nonumber\\&\quad
  \begin{bmatrix}
    \delta x_1\\
    \delta x_2\\
    \vdots\\
    \delta x_N
  \end{bmatrix},
\end{align}
where $f'$ is the derivative of $f$,
and the Jacobian is $DF_1^{-1}=\left|{\mathrm{d} {\xi}}/{ \mathrm{d}x}\right|=1$.

It is also discretised with the trapezoidal scheme (with the drift term at
the midpoint) as
\begin{align}
\hack{\hbox\bgroup\fontsize{8.5}{8.5}\selectfont$\displaystyle}
x_n=x_{n-1}+\frac{f(x_n)+f(x_{n-1})}{2}\delta_t + {\xi}_{n-1},\quad
  n=1,2,\cdots,N,\hack{$\egroup} \label{disc_x2}
\end{align}
which defines a mapping $F_2: {\xi} \mapsto x$. The inverse of the mapping is
linearised as
\begin{align}
& \begin{bmatrix}
    \delta {\xi}_0\\
    \delta {\xi}_1\\
    \vdots\\
    \delta {\xi}_{N-1}
  \end{bmatrix}
=\nonumber\\&\quad
 \hack{\hbox\bgroup\fontsize{7.3}{7.3}\selectfont$\displaystyle} \begin{bmatrix}
    1-\frac{\delta_t}{2}f'(x_1)                        & 0&\cdots &                             0 & 0 \\
    -1-\frac{\delta_t}{2}f'(x_1)& 1-\frac{\delta_t}{2}f'(x_2)&       &                             0 & 0 \\
    \vdots                   &  &       &                               &  \vdots \\
    0                        & 0&\cdots & -1-\frac{\delta_t}{2}f'(x_{N-1}) & 1-\frac{\delta_t}{2}f'(x_{N})
  \end{bmatrix}\hack{$\egroup}\nonumber\\&\quad
  \begin{bmatrix}
    \delta x_1\\
    \delta x_2\\
    \vdots\\
    \delta x_N
  \end{bmatrix},
\end{align}
whose Jacobian is $ DF_2^{-1}=\left|{\mathrm{d} {\xi}}/{\mathrm{d}x}\right| = \prod_{n=1}^N
\left[ 1-(\delta_t/2)f'(x_n) \right] \includegraphics{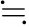} \exp{
  \left[ -(\delta_t/2)\sum_{n=1}^N f'(x_n) \right]
}.
$

\hack{\newpage} Generally, we can assign a measure $\mu_0$ to a cylinder set
${\hat{\Omega}}\equiv \hat{\Omega}_0\times \hat{\Omega}_1\times \cdots \times
\hat{\Omega}_{N-1}$ in the noise space using a density $g$ as follows.
\begin{align}
  \mu_0(\hat{\Omega})&=
  \int_{\hat{\Omega}_0} \mathrm{d} {\xi}_0
  \int_{\hat{\Omega}_1} \mathrm{d} {\xi}_1
  \cdots
  \int_{\hat{\Omega}_{N-1}} \mathrm{d} {\xi}_{N-1}
  g({\xi}_0,{\xi}_1,\cdots,{\xi}_{N-1})\nonumber\\& =\int_{\hat{\Omega}}
  g(\xi) \lambda(\mathrm{d} {\xi})
  =\int_{\hat{\Omega}} \mu_0(\mathrm{d} {\xi}),
\end{align}
where $\lambda$ is the Lebesgue measure on $\mathbb{R}^N$. In our case, we
can see that a small area $\mathrm{d} {\xi}$ in the noise space is equipped with a
measure:
\begin{align}
  \hack{\hbox\bgroup\fontsize{8.5}{8.5}\selectfont$\displaystyle}\mu_0(\mathrm{d} {\xi}) = g({\xi})
  \lambda(\mathrm{d} {\xi}), \quad g({\xi})
  \equiv \frac{1}{(2\pi \delta_t)^{N/2}}
  \mathrm{e}^{-\frac{1}{2\delta_t} \sum_{n=1}^{N}
    \xi_{n-1}^2}.\hack{$\egroup}
\end{align}

Suppose we have a cylinder set $\Omega\equiv \Omega_1\times \Omega_2\times
\cdots \times \Omega_{N}$ in the state space, where each $\Omega_n \subset
\mathbb{R}^1$ is on time slice $t=n\delta_t$. Now, the mapping $F_1$ (or
$F_2$) induces a measure through the change of variables from ${\xi}$ to $x$
with respect to the measure $\mu_0$ as
\begin{align}
  \mu_i(\Omega)&=
  \int_{\Omega_1} \mathrm{d}x_1
  \int_{\Omega_2} \mathrm{d}x_2
  \cdots
  \int_{\Omega_{N}} \mathrm{d}x_{N}
\nonumber\\& (g\circ F_i^{-1})(x_1,x_2,\cdots,x_N) DF_i^{-1}
 \nonumber\\&  =\int_{\Omega}\mu_i(\mathrm{d}x),\quad i=1,2.
\end{align}
In our case, each mapping assigns the following measure to a small area $\mathrm{d}x$
in the corresponding state space:
\begin{align}
  \mu_1(\mathrm{d}x) &\equiv g(F_1^{-1}(x)) DF_1^{-1}  \lambda(  \mathrm{d}x)
  =\nonumber\\&
  \frac{1}{(2\pi \delta_t)^{N/2}}\mathrm{e}^{-\frac{\delta_t}{2}
    \sum_{n=1}^{N} \left(\frac{x_n-x_{n-1}}{\delta_t}-f(x_{n-1})\right)^2}\lambda(\mathrm{d}x),\label{mu1}\\
  \mu_2(\mathrm{d}x) &\equiv g(F_2^{-1}(x)) DF_2^{-1} \lambda(  \mathrm{d}x)
  =\nonumber\\&
  \frac{1}{(2\pi \delta_t)^{N/2}}\mathrm{e}^{-\frac{\delta_t}{2}
    \sum_{n=1}^{N} \left[\left(\frac{x_n-x_{n-1}}{\delta_t}-f(x_{n-\frac12})\right)^2+f'(x_{n})\right]}\nonumber\\&\lambda(\mathrm{d}x),\label{mu2}
\end{align}
where $f(x_{n-\frac12})=\frac{f(x_n)+f(x_{n-1})}{2}.$

Measures $\mu_1$ and $\mu_2$ represent the occurrence probability of the
noise seen from the state space, and thus can be used for path sampling.

The change-of-measure argument (Appendix~\ref{div_in_tra}) or the path
integral argument \citep[e.g.][]{zinn2002quantum} shows that similar forms
are available for time-continuous and multi-dimensional processes, except
that the term $f'(x_t)$ is promoted to $\mathrm{div}\, f(x_t)$.

\subsection{OM functional for mode estimate}\label{OM_MAP}

If we perform path sampling with a sufficient number of paths, in theory we
can find the mean of distribution by averaging the samples, or the mode of
distribution by organising them into a histogram. Still, in some practical
applications, we must efficiently find the mode of distribution by
variational methods; computationally, this approach is much cheaper than path
sampling. For that purpose, we are tempted to use a quadratic cost function
for the minimisation. However, we can illustrate a simple example against
maximising the path probability (\ref{mu1}) to obtain the mode of
distribution. Suppose we have a discrete-time stochastic system in
$\mathbb{R}^1$, starting from $x_0=0$, and we move forward two time steps,
\begin{align}
  x_1&=x_0 + x_0^2 \delta_t + \xi_0= \xi_0,\nonumber\\
  x_2&=x_1 + x_1^2 \delta_t + \xi_1 = \xi_0 + \xi_0^2 \delta_t + \xi_1,
\end{align}
where $\xi_0$ and $\xi_1$ obey independent normal distributions
$\mathcal{N}(0,\delta_t)$. It may be seen as a discrete version of
$\mathrm{d}x_t=x_t^2 \mathrm{d}t+\mathrm{d}w_t$. It is easy to notice that the mode of distribution
$(x_1,x_2)$ is not $(0,0)$ owing to the non-linear term $\xi_0^2 \delta_t$.
On the other hand, according to the path probability (\ref{mu1}),
\begin{align*}
  \mu_1(\mathrm{d}x_1 \mathrm{d}x_2)
  &\propto
  \exp\left[-\frac{\delta_t}{2}
      \left(
      \left(
      \frac{x_1-x_0}{\delta_t}-x_0^2 \right)^2
      \right.\right.\nonumber\\&+\left.\left.
      \left(\frac{x_2-x_1}{\delta_t}-x_1^2 \right)^2
      \right)\right]\lambda(\mathrm{d}x_1 \mathrm{d}x_2),
\end{align*}
the best trajectory is $(x_1,x_2)=(0,0)$, which has no noise:
$(\xi_0,\xi_1)=(0,0).$ We expect a path with the highest probability at
$(x_1,x_2)=(0,0)$, but it is not the route where the paths are most
concentrated.

Motivated by this example, we shall investigate a proper strategy to find the
route that maximises the density of paths. In this regard, we ask how densely
the paths populate in the small neighbourhood of a curve $\phi=\phi(t)$ in
the state space.

Assuming that $f$ and $\phi$ are twice continuously differentiable, we
evaluate the density of paths in the $\epsilon$-neighbourhoods around a curve
$\phi$ connecting points $\{\phi_n, ~ n=1,2,\cdots,N \}$ with the following
integral:
  \begin{align}
    I_{\epsilon,\delta_t}(\phi)
    &=
    \int_{\phi_1-\epsilon}^{\phi_1+\epsilon}\mathrm{d}x_1
    \int_{\phi_2-\epsilon}^{\phi_2+\epsilon}\mathrm{d}x_2
    \cdots \int_{\phi_N-\epsilon}^{\phi_N+\epsilon}\mathrm{d}x_N
    \frac{1}{(2 \pi \delta_t)^{N/2}}
   \nonumber\\& \exp{\left\{
      -\frac{\delta_t}{2}
      \sum_{n=1}^N \left(
      \frac{x_n-x_{n-1}}{\delta_t}-f(x_{n-1})
      \right)^2
      \right\}}\label{I_1}\\
    &=
    \int_{-\epsilon}^{\epsilon}\mathrm{d}v_1
    \int_{-\epsilon}^{\epsilon}\mathrm{d}v_2
    \cdots \int_{-\epsilon}^{\epsilon}\mathrm{d}v_N
    \frac{1}{(2 \pi \delta_t)^{N/2}}
   \nonumber\\& \exp\left\{
      -\frac{\delta_t}{2}
      \sum_{n=1}^N \left(
      \frac{v_n-v_{n-1}}{\delta_t}
      +\frac{\phi_n-\phi_{n-1}}{\delta_t}
    \right. \right.\nonumber\\& \quad \left.\left. -f(v_{n-1}+\phi_{n-1})
      \right)^2
      \right\}\label{I_2}\\
    &=
    \int_{-\epsilon}^{\epsilon}\mathrm{d}v_1
    \int_{-\epsilon}^{\epsilon}\mathrm{d}v_2
    \cdots \int_{-\epsilon}^{\epsilon}\mathrm{d}v_N
    \frac{1}{(2 \pi \delta_t)^{N/2}}
 \nonumber\\&   \exp{\left\{
      -\frac{\delta_t}{2}
      \sum_{n=1}^N \left(
      \frac{v_n-v_{n-1}}{\delta_t}
      \right)^2
      \right\}}\nonumber
    \\
    &\times
    \exp\left\{
      -\frac{\delta_t}{2}
      \sum_{n=1}^N
      \left[
        \left
        (\frac{\phi_n-\phi_{n-1}}{\delta_t}
        -f(v_{n-1}+\phi_{n-1})
        \right)^2
     \right.\right.\nonumber\\&\quad \left.\left.   +2
        \left(\frac{\phi_n-\phi_{n-1}}{\delta_t}
        -f(v_{n-1}+\phi_{n-1})
        \right)
       \right.\right.\nonumber\\&\quad\left.\left. \left(
        \frac{v_n-v_{n-1}}{\delta_t}
        \right)
        \right]
      \right\}.\label{I_dv_ens}
  \end{align}
By regarding $v_n$ in Eq.~(\ref{I_dv_ens}) as being generated according to
the probability $\frac{1}{(2 \pi \delta_t)^{N/2}}\mathrm{e}^{
  -\frac{\delta_t}{2}
  \sum_{n=1}^N \left(
  \frac{v_n-v_{n-1}}{\delta_t}
  \right)^2
}$, we can interpret the integration as a weighted ensemble averaging of a
random function up to a numerical constant. The sequence $v_n$ can be set as
a random walk $v_0=0,~v_n= \sum_{k=1}^n \xi_k$, where $\xi_k$ are independent
normal random variables obeying $\mathcal{N}(0,\delta_t)$. For simplicity, we
rather assume that $\xi_k$ takes values $\pm \sqrt{\delta_t}$ with $0.5$
probability for either one, because Donsker's theorem ensures it has the same
probability law as the former when $\delta_t$ is sufficiently small. We
suppose $\sqrt{\delta_t}<\epsilon$ so that no step of the random walk escapes
from the $\epsilon$-neighbourhood. Accordingly, the integral is expressed as
the ensemble average with respect to random walks confined in the tube
$[0,N\delta_t] \times [-\epsilon,\epsilon]$:
\begin{align}
  I_{\epsilon,\delta_t}(\phi)
  &\propto \mathbb{E}_v\left[\mathrm{e}^{-J(\phi,v)} \Big|
    (\forall n)~\left|v_n\right|<\epsilon \right],\label{eq_I}\\
  J(\phi,v) &\equiv
  \frac{\delta_t}{2}
  \sum_{n=1}^N
  \left[
    \left
    (\frac{\phi_n-\phi_{n-1}}{\delta_t}
    -f(v_{n-1}+\phi_{n-1})
    \right)^2
   \right.\nonumber\\&\quad \left. +2
    \left(\frac{\phi_n-\phi_{n-1}}{\delta_t}
    -f(v_{n-1}+\phi_{n-1})
    \right)
   \right.\nonumber\\&\quad\left. \left(
    \frac{v_n-v_{n-1}}{\delta_t}
    \right)
    \right],
\end{align}
where $\mathbb{E}_v$ denotes the ensemble averaging of the random walks
denoted by $v$, each of which follows the route $(v_0,v_1,\cdots,v_N)$ and
satisfies $|v_n|<\epsilon$ for all $n$.

Because $v_{n-1}$ is small, we can apply the expansion
\begin{align}
  f(v_{n-1}+\phi_{n-1})
  =f(\phi_{n-1})+f'(\phi_{n-1})v_{n-1}+O(v^2),
\end{align}
where $f'$ is the derivative of $f$. Let us accept that the following average
containing the higher-order terms $O(v^2)$ converges (see Eq.~\ref{conv_1}).
\begin{align}
  \mathbb{E}_v\left[
    \mathrm{e}^{ \sum_{n=1}^N O(v^2) (v_n-v_{n-1})}
 \Big|
    (\forall n)~\left|v_n\right|<\epsilon
    \right] &
  \xrightarrow{\epsilon\to 0}
  1. \label{conv_v2}
\end{align}
As shown in Appendix~\ref{Girsa}, the remaining terms in the exponent
$-J(\phi,v)$ are less than $O(\epsilon)$, except for the following one.
\begin{align}
  &\sum_{n=1}^N
  f'(\phi_{n-1})
  v_{n-1}\left(v_n-v_{n-1}\right)
  =
  \sum_{n=1}^N
  f'(\phi_{n-1})
\nonumber\\&\quad \left[
    \frac12(v_{n-1}-v_{n})
    +
    \frac12(v_{n-1}+v_{n})
    \right]\left(v_n-v_{n-1}\right)\\
  &=
  \sum_{n=1}^N
  f'(\phi_{n-1})
  \frac12(v_{n-1}-v_{n})\left(v_n-v_{n-1}\right)
 \nonumber\\&\quad   +
  \sum_{n=1}^N
  f'(\phi_{n-1})
  \frac12(v_{n}^2-v_{n-1}^2)\\
  &=
  -\frac12 \sum_{n=1}^N
  f'(\phi_{n-1})
  \xi_n^2
  +
  \frac12 \sum_{n=1}^{N-1}
  \left[ f'(\phi(t_{n-1}))\right. \nonumber\\&\quad \left.-f'(\phi(t_{n-1}+\delta_t)) \right]
  v_n^2
  +
  \frac12 f'(\phi_{N-1}) v_N^2\\
  &= -\frac{\delta_t}2 \sum_{n=1}^N
  f'(\phi_{n-1})+O(\epsilon^2). \nonumber\\&\quad \includegraphics{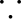} \xi_n=\pm
  \sqrt{\delta_t},\quad
  f'(\phi(t_{n-1}))-f'(\phi(t_{n-1}+\delta_t))=O(\delta_t), \nonumber\\&\quad  v_n^2<\epsilon^2.
\end{align}

Consequently, we obtain the asymptotic expression for the ensemble average
when $\epsilon$ is small and $\delta_t<\epsilon^2$:
\begin{align}
& I_{\epsilon,\delta_t}(\phi)
 \propto
  \mathbb{E}_v    \nonumber\\&\quad \hack{\hbox\bgroup\fontsize{8.5}{8.5}\selectfont$\displaystyle} \left[
    \mathrm{e}^{
      -\frac{\delta_t}2  \sum_{n=1}^N
      \left[
        \left(
        \frac{\phi_n-\phi_{n-1}}{\delta_t}-f(\phi_{n-1})
        \right)^2
        +f'(\phi_{n-1})
        \right]
  +O(\epsilon)
      +\sum_{n=1}^N O(v^2) (v_n-v_{n-1})
    }\right.   \hack{$\egroup} \nonumber\\&\quad \left.
 \Big|
    (\forall n)~\left|v_n\right|<\epsilon  \right]
  \\
  &\rightarrow
  \mathrm{e}^{
    -\frac12 \int_0^T
    \left[ \left(\dot{\phi}(t) -f(\phi(t)) \right)^2
      +f'(\phi(t)) \right] \mathrm{d}t }.\label{I2OM}
\end{align}
Appendix~\ref{Girsa} shows that a similar form is available for
time-continuous and multi-dimensional processes, except that the term
$f'(\phi(t))$ is promoted to $\mathrm{div}\, f(\phi(t))$.

Importantly, the control variable for the optimisation has changed from $x$
to $\phi$.

\subsection{Probabilistic description of data assimilation}\label{sec_prob}

Using the OM functional derived in Sect.~\ref{OM_path} and \ref{OM_MAP} as a
model error term, we shall develop a probabilistic description of data
assimilation.

Following the derivation in Sect.~2.3 of \citet{law2015data2}, we can assign
each path a posterior probability
\begin{align}
  P(x|y)&\propto P(x) P(y|x)
  =
  P(x|x_0) P(x_0) P(y|x)
 \nonumber\\& =
  \prod_{n=1}^{N} P(x_{n}|x_{n-1}) P(x_0)
  \prod_{m\in M} P(y_m|x_m).
  \label{Pxy}
\end{align}
According to Eq.~(\ref{SDE1bg}), the prior probability for the initial
condition is given as
\begin{align}
  P(x_0) &\propto \exp{\left(-
    \frac{|x_0-x_\mathrm{b}|^2}{2\sigma_\mathrm{b}^2}
    \right)},\label{P_b}
\end{align}
where $|x_0-x_\mathrm{b}|^2$ represents the squared Euclidean norm
$\sum_{i=1}^{D} (x^i_0-x^i_\mathrm{b})^2$. According to Eq.~(\ref{SDE1obs}),
the likelihood of the state $x_m$, given observation $y_m$, is
\begin{align}
  P(y_m|x_m) &\propto \exp{\left(-
    \frac{|y_m-x_m|^2}{2\sigma_\mathrm{o}^2}
    \right)}.\label{P_o}
\end{align}

Based on the argument in Sect.~\ref{OM_path}, Eq.~(\ref{disc_x}) has the
transition probability at discrete time steps
\begin{align}
  P(x_{n}|x_{n-1})&\propto
  \exp{
    \left(-
    \frac{\delta_t}{2\sigma^2} \left| \frac{x_{n}-x_{n-1}}{\delta_t}-f(x_{n-1})\right|^2
    \right)},
\end{align}
called the Euler scheme, which uses the drift $f(x_{n-1})$ at the previous
time step. Section~\ref{OM_path} also shows that this transition probability
has another expression:
\begin{align}
& P(x_{n}|x_{n-1})\propto
 \nonumber\\&\quad  \hack{\hbox\bgroup\fontsize{8.5}{8.5}\selectfont$\displaystyle}\exp{
    \left(-
    \frac{\delta_t}{2\sigma^2} \left| \frac{x_{n}-x_{n-1}}{\delta_t}-f(x_{n-\frac12})\right|^2-\frac{\delta_t}{2}\mathrm{div}\,{f}(x_{n})
    \right)},\hack{$\egroup}\\
  &f(x_{n-\frac12})\equiv \frac{f(x_n)+f(x_{n-1})}2, \nonumber\\&
  \quad \mathrm{div}\,{f}(x) \equiv \sum_{i=1}^D \frac{\partial f^i}{\partial x^i}(x),
\end{align}
which can be called the trapezoidal scheme because the integral is evaluated
with the drift terms at both ends of each interval. The transition
probability leads to the prior probability $P(x|x_0)$ of a path
$x=\{x_n\}_{0\le n \le N}$ as follows:
\begin{align}
  P(x|x_0)
  &\hack{\hbox\bgroup\fontsize{8.5}{8.5}\selectfont$\displaystyle}\propto
  \exp{
    \left(-\delta_t
    \sum_{n=1}^N
    \frac{1}{2\sigma^2} \left| \frac{x_{n}-x_{n-1}}{\delta_t}-f(x_{n-1})\right|^2
    \right)}\hack{$\egroup}\\
  &\leftrightharpoons
  \exp
    \left(-\delta_t
    \sum_{n=1}^N \left[
      \frac{1}{2\sigma^2} \left|
      \frac{x_{n}-x_{n-1}}{\delta_t}-f(x_{n-\frac12})\right|^2
      \right.\right.\nonumber\\&\quad\left.\left.+\frac12\mathrm{div}\,{f}(x_{n})\right]
    \right),
\end{align}
where the ``$\leftrightharpoons$'' sign indicates that, if $\delta_t$ is
sufficiently small, the equations on both sides are compatible.

On the other hand, based on the argument in Sect.~\ref{OM_MAP}, we can also
define the probability $P(U_{\phi}|\phi_0)$ for a smooth tube that represents
its neighbouring paths $U_{\phi} = \left\{ \omega \Big| (\forall n)
|\phi_n-x_n(\omega)|<\epsilon \right\}$:
\begin{align}
  P(U_{\phi}|\phi_0) &\propto
  \exp
    \left(-\delta_t
    \sum_{n=1}^N \left[
      \frac{1}{2\sigma^2} \left| \frac{\phi_{n}-\phi_{n-1}}{\delta_t}-f(\phi_{n-1})\right|^2
    \right.\right.\nonumber\\&\quad \left.\left.  +
      \frac{1}2 \mathrm{div}\,{f(\phi_{n-1})}
      \right]
    \right)
  .\label{PI_ED}
\end{align}
The scaling argument for a smooth curve in Appendix~\ref{scale} allows us to
use the drift term $f(\phi_{n-\frac12})$ instead in Eq.~(\ref{PI_ED}):
\begin{align}
  P(U_{\phi}|\phi_0) &\propto
  \exp
    \left(-\delta_t
    \sum_{n=1}^N \left[
      \frac{1}{2\sigma^2} \left| \frac{\phi_{n}-\phi_{n-1}}{\delta_t}-f(\phi_{n-\frac12})\right|^2
     \right.\right.\nonumber\\&\quad  \left.\left.+
      \frac{1}2 \mathrm{div}\,{f(\phi_{n-\frac12})}
      \right]
    \right)
  .\label{PI_TD}
\end{align}

The corresponding posterior probabilities are thus given as follows:
\begin{align}
 & P_{\text{path}}(x|y) \propto \exp{(-J_{\text{path}}(x|y))},\\
 & J_{\text{path}}(x|y)\equiv
  \frac1{2\sigma_\mathrm{b}^2}\left|x_0-x_\mathrm{b}\right|^2
 \nonumber\\& \quad +
  \sum_{m\in M} \frac1{2\sigma_\mathrm{o}^2}\left|x_m-y_m\right|^2
  +
  \delta_t \sum_{n=1}^N
  \nonumber\\& \left(
  \frac{1}{2\sigma^2}\left|
  \frac{x_n-x_{n-1}}{\delta_t}-f(x_{n-1})
  \right|^2
  \right)\label{Jp_E}\\
  &\leftrightharpoons
  \frac1{2\sigma_\mathrm{b}^2}\left|x_0-x_\mathrm{b}\right|^2
  +
  \sum_{m\in M} \frac1{2\sigma_\mathrm{o}^2}\left|x_m-y_m\right|^2
 \nonumber\\& \quad \hack{\hbox\bgroup\fontsize{8.5}{8.5}\selectfont$\displaystyle} +
  \delta_t \sum_{n=1}^N
  \left(
  \frac{1}{2\sigma^2}\left|
  \frac{x_n-x_{n-1}}{\delta_t}-f(x_{n-\frac12})
  \right|^2
  +
  \frac{1}{2}
  \mathrm{div}\, f(x_{n})
  \right)\hack{$\egroup}\label{Jp_TD}
\end{align}
for a sample path, and
\begin{align}
& P_{\text{tube}}(U_{\phi}|y) \propto
  P(U_{\phi}|\phi_0)P(\phi_0)P(y|U_{\phi})
  \propto
 \nonumber\\& \quad  \exp{(-J_{\text{tube}}(\phi|y))},\\
& J_{\text{tube}}(\phi|y)\equiv
  \frac1{2\sigma_\mathrm{b}^2}\left|\phi_0-x_\mathrm{b}\right|^2
  +
  \sum_{m\in M} \frac1{2\sigma_\mathrm{o}^2}\left|\phi_m-y_m\right|^2
 \nonumber\\& \quad  \hack{\hbox\bgroup\fontsize{8.5}{8.5}\selectfont$\displaystyle} +
  \delta_t \sum_{n=1}^N
  \left(
  \frac{1}{2\sigma^2}\left|
  \frac{\phi_n-\phi_{n-1}}{\delta_t}-f(\phi_{n-\frac12})
  \right|^2
  +
  \frac{1}{2}
  \mathrm{div}\, f(\phi_{n-\frac12})
  \right)\hack{$\egroup}\label{Jt_TD}\\
  &\leftrightharpoons
  \frac1{2\sigma_\mathrm{b}^2}\left|\phi_0-x_\mathrm{b}\right|^2
  +
  \sum_{m\in M} \frac1{2\sigma_\mathrm{o}^2}\left|\phi_m-y_m\right|^2
 \nonumber\\& \quad \hack{\hbox\bgroup\fontsize{8.5}{8.5}\selectfont$\displaystyle}  +
  \delta_t \sum_{n=1}^N
  \left(
  \frac{1}{2\sigma^2}\left|
  \frac{\phi_n-\phi_{n-1}}{\delta_t}-f(\phi_{n-1})
  \right|^2
  +
  \frac{1}{2}
  \mathrm{div}\, f(\phi_{n-1})
  \right),\hack{$\egroup}\label{Jt_ED}
\end{align}
for a smooth tube. Note that different pairs of time-discretisation schemes
of the OM functional, $\frac1{2\sigma^2} \left( \frac{\mathrm{d}x}{\mathrm{d}t}-f(x)\right)^2
+\frac12 \mathrm{div}\,{(f)}$, are nominated for paths and for tubes in
Eqs.~(\ref{Jp_E}), (\ref{Jp_TD}), (\ref{Jt_TD}), and (\ref{Jt_ED}).

\section{Method}\label{sec_method}
\subsection{Four schemes for OM}\label{OMschemes}

In the argument in Sects.~\ref{OM_path} and \ref{OM_MAP}, the prior
probability has a form $P(x|x_0)\propto\exp{\left(-\delta_t\sum_{n=1}^N
\widetilde{\mathrm{OM}} \right)},$ where $\widetilde{\mathrm{OM}}$ is the OM
functional. As a proof-of-concept described in these sections, we will test
all the cases with conceivable combinations of the timing of the drift term
$f(x_t)$ and the presence or absence of the divergence term. Including those
shown in Eqs.~(\ref{Jp_E}), (\ref{Jp_TD}), (\ref{Jt_TD}), and (\ref{Jt_ED}),
as well as those that are potentially incorrect, the possible candidates for
the discretisation schemes of the OM functional are as follows, where the
symbol $\psi$ represents either $\phi$ for a smooth curve or $x$ for a sample
path.
\begin{enumerate}
\item Euler scheme (E) \citep[e.g.][]{zinn2002quantum,dutra2014maximum}:
  \begin{align}
   & \widetilde{\mathrm{OM}}_{\text{E}} \equiv \frac1{2\sigma^2} \left|
    \frac{\psi_{n}-\psi_{n-1}}{\delta_t}-f{(\psi_{n-1})}\right|^2;
  \end{align}
\item Euler scheme with divergence term (ED):
  \begin{align}
   & \widetilde{\mathrm{OM}}_{\text{ED}} \equiv \frac1{2\sigma^2} \left|
    \frac{\psi_{n}-\psi_{n-1}}{\delta_t}-f{(\psi_{n-1})}\right|^2 \nonumber\\&\quad + \frac12
    \mathrm{div}\, f{(\psi_{n-1})};
  \end{align}
\item trapezoidal scheme (T):
  \begin{align}
   & \widetilde{\mathrm{OM}}_{\text{T}} \equiv
    \frac1{2\sigma^2} \left| \frac{\psi_{n}-\psi_{n-1}}{\delta_t}
    -f{(\psi_{n-\frac12})}\right|^2;
  \end{align}
\item trapezoidal scheme with divergence term (TD) \citep[e.g.][]{ikeda2014stochastic,ISI:000247587700006,dutra2014maximum}:
  \begin{align}
   & \widetilde{\mathrm{OM}}_{\text{TD}} \equiv
    \frac1{2\sigma^2} \left|
    \frac{\psi_{n}-\psi_{n-1}}{\delta_t}-f{(\psi_{n-\frac12})}\right|^2\nonumber\\&\quad  +
    \frac12 \mathrm{div}\, f{(\psi_{n-\frac12})},
  \end{align}
  where $f(\psi_{n-\frac12}) = (f(\psi_n)+f(\psi_{n-1}))/2$.
\end{enumerate}

\subsection{Data assimilation algorithms}

By using one of the above schemes adopted for the model error term in the
cost function, we can apply a data assimilation algorithm -- either Markov
chain Monte Carlo (MCMC) \citep[e.g.][]{metropolis1953equation} or
four-dimensional variational data assimilation (4D-Var)
\citep[e.g.][]{zupanski1997general}. Among versions of MCMC, we focus on the
Metropolis-adjusted Langevin algorithm (MALA)
\citep[e.g.][]{roberts1998optimal,cotter2013mcmc}. MALA samples the paths
$x^{(k)}=\{x_n(\omega_k)\}_{0\le n \le N}$ according to the distribution
$P_{\text{path}}$ by iterating
\begin{align}  x^{(k+1)} &= x^{(k)} - \alpha
 \nabla J_{\text{path}} +\sqrt{2\alpha} \xi,
\nonumber\\& \alpha>0,~
  \xi \sim \mathcal{N}(0,1)^{D(N+1)},~
  \nabla J=\left( \frac{\partial J}{\partial x} \right)^T,
\end{align}
with the Metropolis rejection step for adjustment, to obtain an ensemble of
sample paths according to the posterior probability, while 4D-Var seeks the
centre of the most probable tube $\phi=\{\phi_n\}_{0\le n \le N}$ by
iterating:
\begin{align}
 \phi^{(k+1)} &= \phi^{(k)}-\alpha \nabla J_{\text{tube}},
 \quad  \alpha>0.\label{itr_4dvar}
\end{align}
Note that if the OM functional of type $\widetilde{\mathrm{OM}}_{\text{ED}}$
is used, the gradient is of the form
\begin{align}
  \nabla_{\phi_n} J_{\text{tube}}&=
  \frac1{\sigma_\mathrm{b}^2}(\phi_0-x_\mathrm{b})\delta_{0,n}
  +
  \sum_{m\in M} \frac1{\sigma_\mathrm{o}^2}(\phi_m-y_m)\delta_{m,n}
  \nonumber\\
  &+
  \frac{1}{\sigma^2}
  \left(
  \frac{\phi_n-\phi_{n-1}}{\delta_t}-f(\phi_{n-1})
  \right) \qquad \qquad (n>0)\nonumber\\
  &+\frac{\delta_t}{\sigma^2}
  \left(
  -\frac1{\delta_t}-\left(\frac{\partial f}{\partial \phi_n}(\phi_n)\right)^T
  \right)
\nonumber\\&\quad
\left(\frac{\phi_{n+1}-\phi_{n}}{\delta_t}-f(\phi_{n})\right)
  +\frac{\delta_t}{2}\frac{\partial}{\partial \phi_n} \mathrm{div}\,
  f(\phi_n)  \nonumber\\&\quad (n<N),
  \label{gradient_phi}
\end{align}
where $\left(\frac{\partial f}{\partial \phi_n}(\phi_n)\right)^T$ is an
adjoint integration starting from the subsequent term, which is typical
in gradient calculations in 4D-Var.
In comparison, the term $\frac{\partial}{\partial \phi_n} \mathrm{div}\,
f(\phi_n)$ requires the second derivative of $f$, which is not typical in
4D-Var, and could be difficult to implement in large dimensional systems.

To investigate the applicability of the four candidate schemes in
Sect.~\ref{OMschemes}, we use them in these algorithms.

The results should be checked with ``the correct answer''. The reference
solution that approximates the correct answer is provided by a particle
smoother (PS) \citep[e.g.][]{doucet2000sequential}, which does not involve
the explicit computation of prior probability. When we have observations only
at the end of the assimilation window, the PS algorithm is as follows.
\begin{enumerate}
\item Generate samples of initial and
  model errors, integrate $M$ copies of the model, and use them
to obtain a Monte Carlo approximation of the prior distribution:
  \begin{align}
    P(x) &\simeq \frac1M \sum_{m=1}^M \prod_{n=0}^N \delta(x_n-\chi_n^{(m)}),
  \end{align}
  where $\chi_n^{(m)}$ is the state of member $m$ at time $n$.
\item
  Reweight it according to Bayes' theorem:
  \begin{align}
    P(y|x)&\propto \exp{\left(-\frac1{2\sigma_\mathrm{o}^2}
      |y-x_N|^2\right)}, \\
    P(x|y)&=\frac{P(x)P(y|x)}{\int \mathrm{d}x P(x)P(y|x) }
    = \sum_{m=1}^M \prod_{n=0}^N \delta(x_n-\chi_n^{(m)})
\nonumber\\&    \frac{w^{(m)}}{\sum_{m=1}^M w^{(m)}},\\
    w^{(m)}&\equiv \exp{\left(-\frac1{2\sigma_\mathrm{o}^2} |y-\chi^{(m)}_N|^2\right)}.
  \end{align}

\end{enumerate}

\section{Results}
\subsection{Example A (hyperbolic model)}

In our first example, we solve the non-linear smoothing problem for the
hyperbolic model \citep{daum1986exact}, which is a simple problem with
one-dimensional state space, but which has a non-linear drift term. We want
to find the probability distribution of the paths described by
\begin{align}
  \mathrm{d}x_t &= \tanh(x_t) \mathrm{d}t + \mathrm{d}w_t, \quad  x_{t=0} \sim \mathcal{N}(0,0.16),
  \quad
  \label{tanh}
\end{align}
subject to an observation $y$:
\begin{align}
  y |x_{t=5} &\sim \mathcal{N}(x_{t=5},0.16),\quad y=1.5.
\end{align}
The setting follows \citet{daum1986exact}. In this case,
$\mathrm{div}\,{f(x)}=1/\cosh^2(x)$ imposes a penalty for small $x$. The
total time duration $T=5$ is divided into $N=100$ segments with
$\delta_t=5\times 10^{-2}$.

\begin{figure*}[t]
\includegraphics[width=13.5cm]{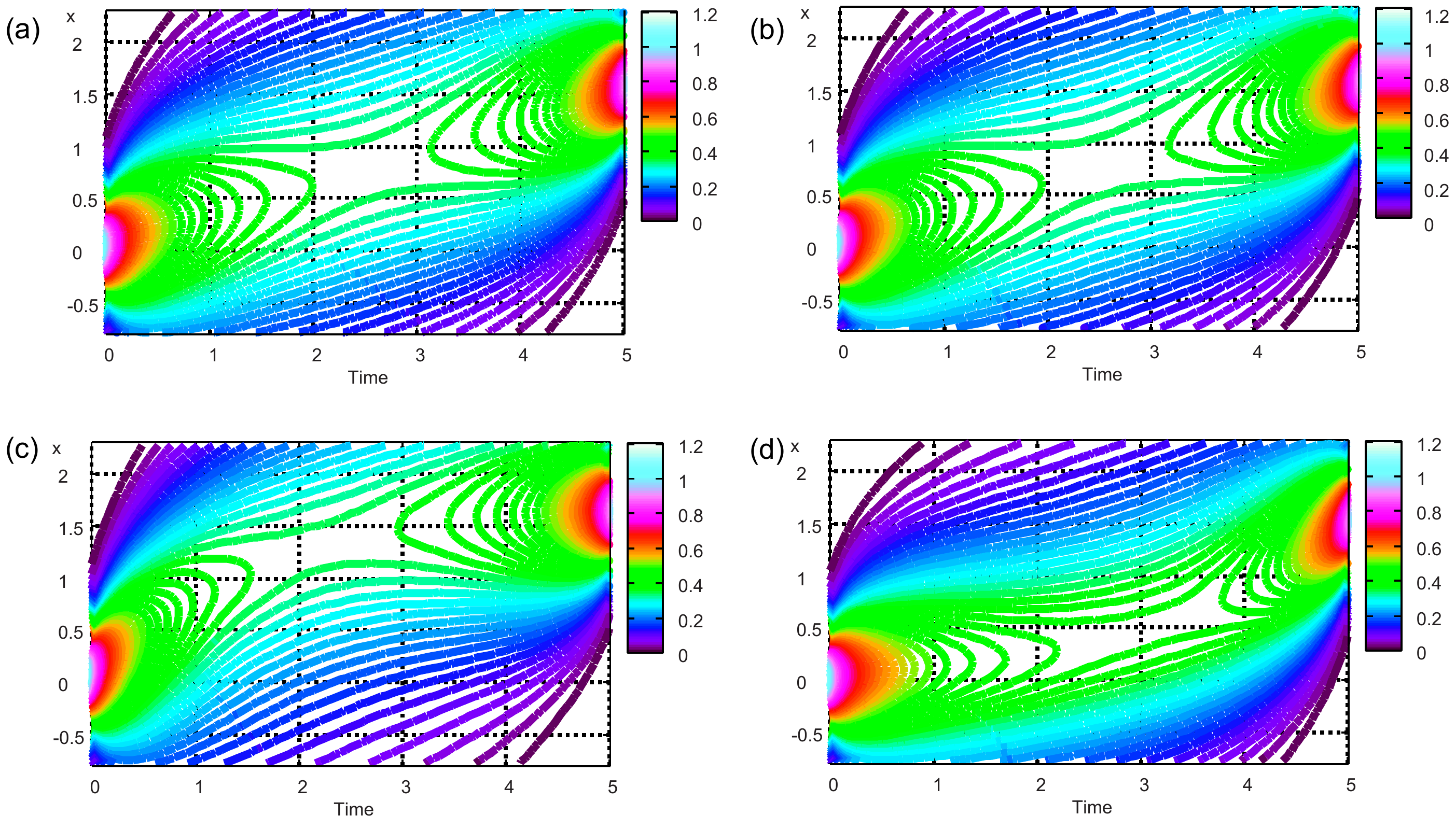}
  \caption{Probability density of paths derived by MCMC and PS for the hyperbolic model. \textbf{(a)}~Reference solution by
  PS, \textbf{(b)}~solution by MCMC with scheme E or TD, \textbf{(c)}~solution by MCMC with scheme ED, and \textbf{(d)}~solution by MCMC with
  scheme T.\label{PD}}
\end{figure*}

\begin{figure}[t]
\includegraphics[width=8.5cm]{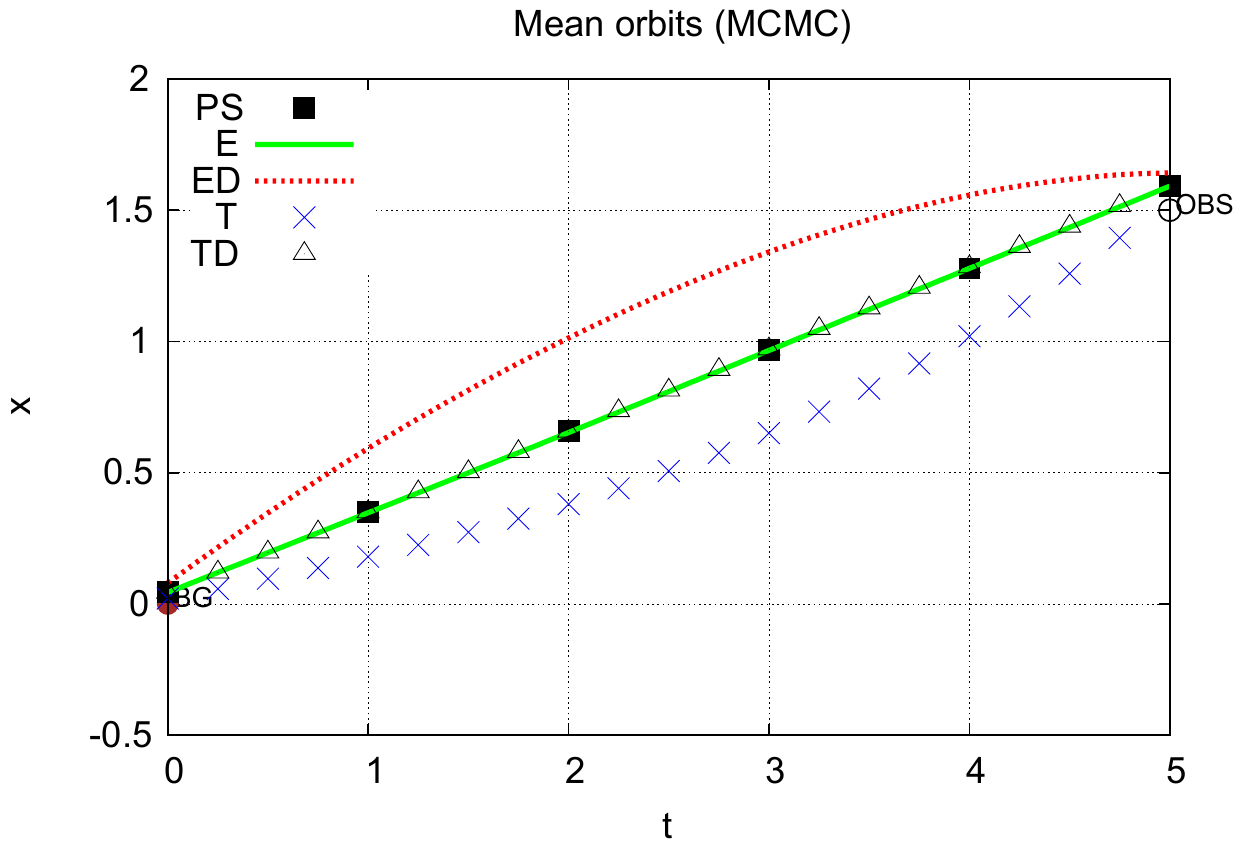}
\caption{Expected path derived by MCMC (hyperbolic model).
    \label{ExpTanh}}
\end{figure}

Figure \ref{PD} shows the probability densities of paths normalised on each
time slice, $P_{t=n}(\phi)=\int P(U_{\phi}|y) \mathrm{d}\phi_{t\neq n}$, derived by
MCMC and PS. PS is performed with $5.1\times 10^{10}$ particles. It is clear
that MCMC with E or TD provides the proper distribution matched with that of
PS; this is also clear from the expected paths yielded by these experiments,
as shown in Fig.~\ref{ExpTanh}. These schemes correspond to candidates in
Eqs.~(\ref{Jp_E}) and (\ref{Jp_TD}). The expected path by ED bends towards a
larger $x$, which should be caused by an extra penalty for a larger $x$. The
expected path by T bends towards a smaller $x$, which should be caused by the
lack of a penalty for a larger $x$.

\begin{figure}[t]
\includegraphics[width=8.5cm]{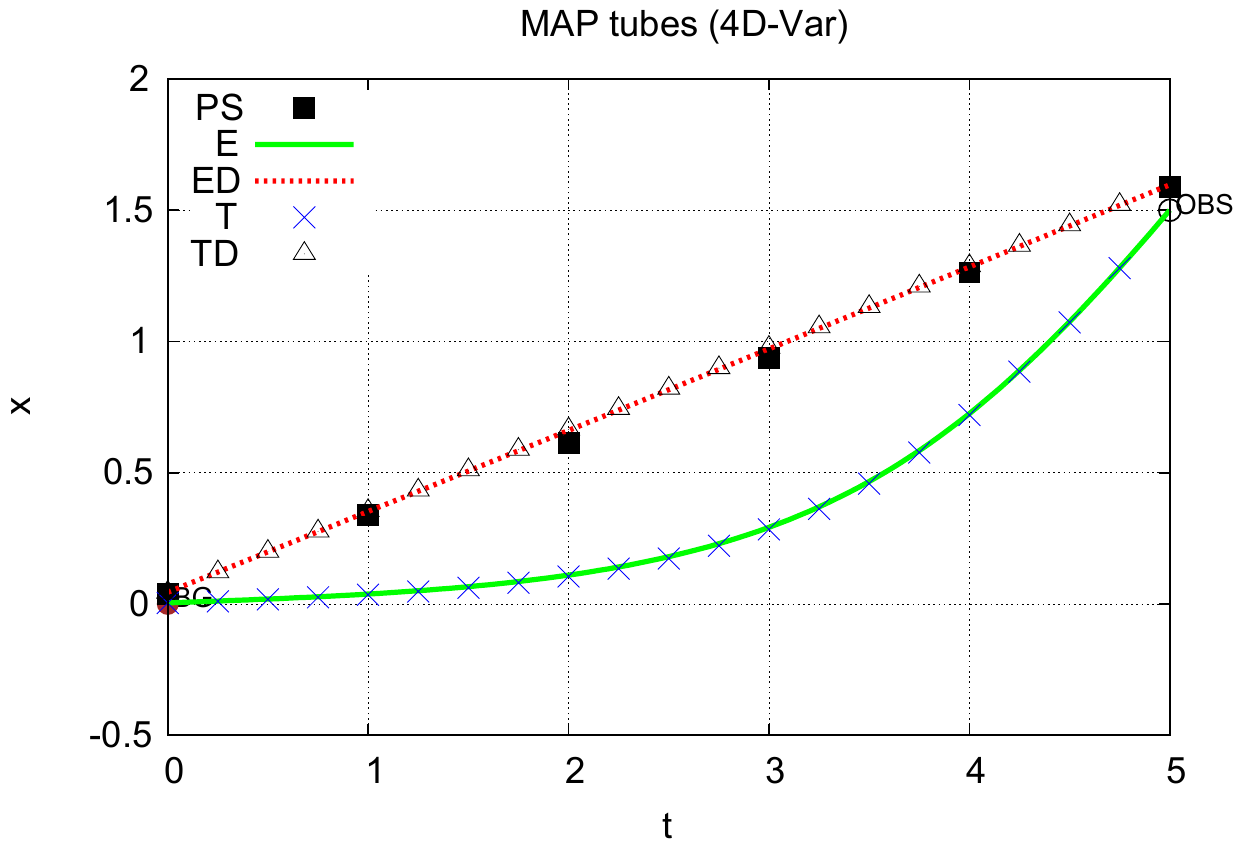}
  \caption{Most probable tube derived by 4D-Var (hyperbolic model).
    \label{MAPTanh}}
\end{figure}

The results of 4D-Var, which represents the MAP estimates, are shown in
Fig.~\ref{MAPTanh}. ED and TD provide the proper MAP estimate. These schemes
correspond to candidates in Eqs.~(\ref{Jt_TD}) and (\ref{Jt_ED}). The
expected paths by E and T bend towards a smaller $\phi$, which should be
caused by the lack of a penalty for a larger $\phi$.

\subsection{Example B (R\"ossler model)}

In our second example, we solve the non-linear smoothing problem for the
stochastic R\"ossler model \citep{ROSSLER1976397}. We want to find the
probability distribution of the paths described by
\begin{align}
  \begin{cases}
    \mathrm{d}x_1 &= (-x_2-x_3) \mathrm{d}t    + \sigma \mathrm{d}w_1,\\
    \mathrm{d}x_2 &= (x_1+a x_2) \mathrm{d}t    + \sigma \mathrm{d}w_2,\\
    \mathrm{d}x_3 &= (b+x_1 x_3-c x_3) \mathrm{d}t + \sigma \mathrm{d}w_3,
  \end{cases}
\end{align}
\begin{align}
  x_{t=0} &\sim \mathcal{N}(x_\mathrm{b},0.04I),
\end{align}
subject to an observation $y$:
\begin{align}
  y |x_{t=0.4} &\sim \mathcal{N}(x_{t=0.4},0.04I),
\end{align}
where $(a, b, c)=(0.2, 0.2, 6),~ \sigma=2,$ $x_\mathrm{b} = ( 2.0659834,
-0.2977757, 2.0526298)^T,$ and $y = ( 2.5597086, 0.5412736, 0.6110939)^T$. In
this case, $\mathrm{div}\,{f(x)}=x_1+a-c$ imposes a penalty for large $x_1$.
The total time duration $T=0.4$ is divided into $N=800$ segments with
$\delta_t=5\times 10^{-4}$.

\begin{figure}[t]
\includegraphics[width=8.5cm]{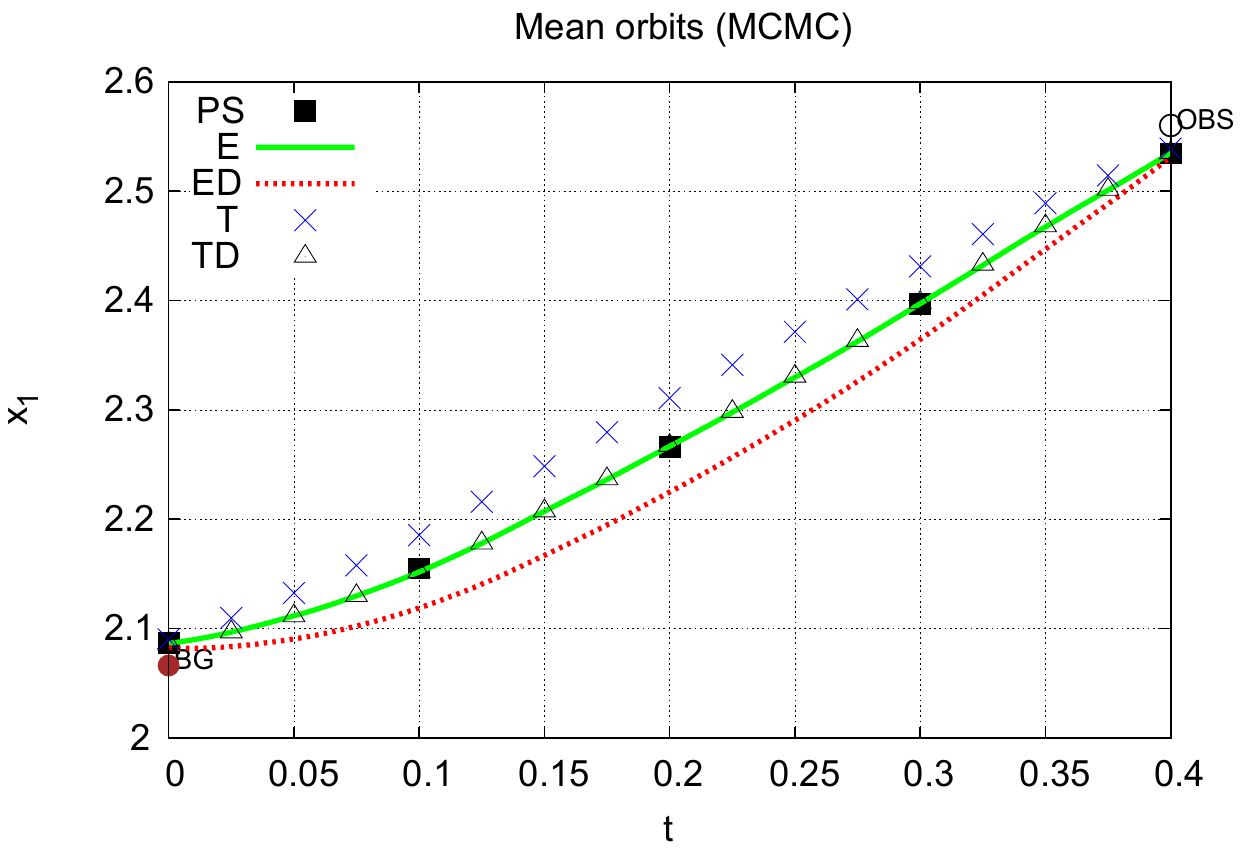}
  \caption{Expected path derived by MCMC (R\"ossler model).
    \label{ExpRoss}}
\end{figure}

\begin{figure}[t]
\includegraphics[width=8.5cm]{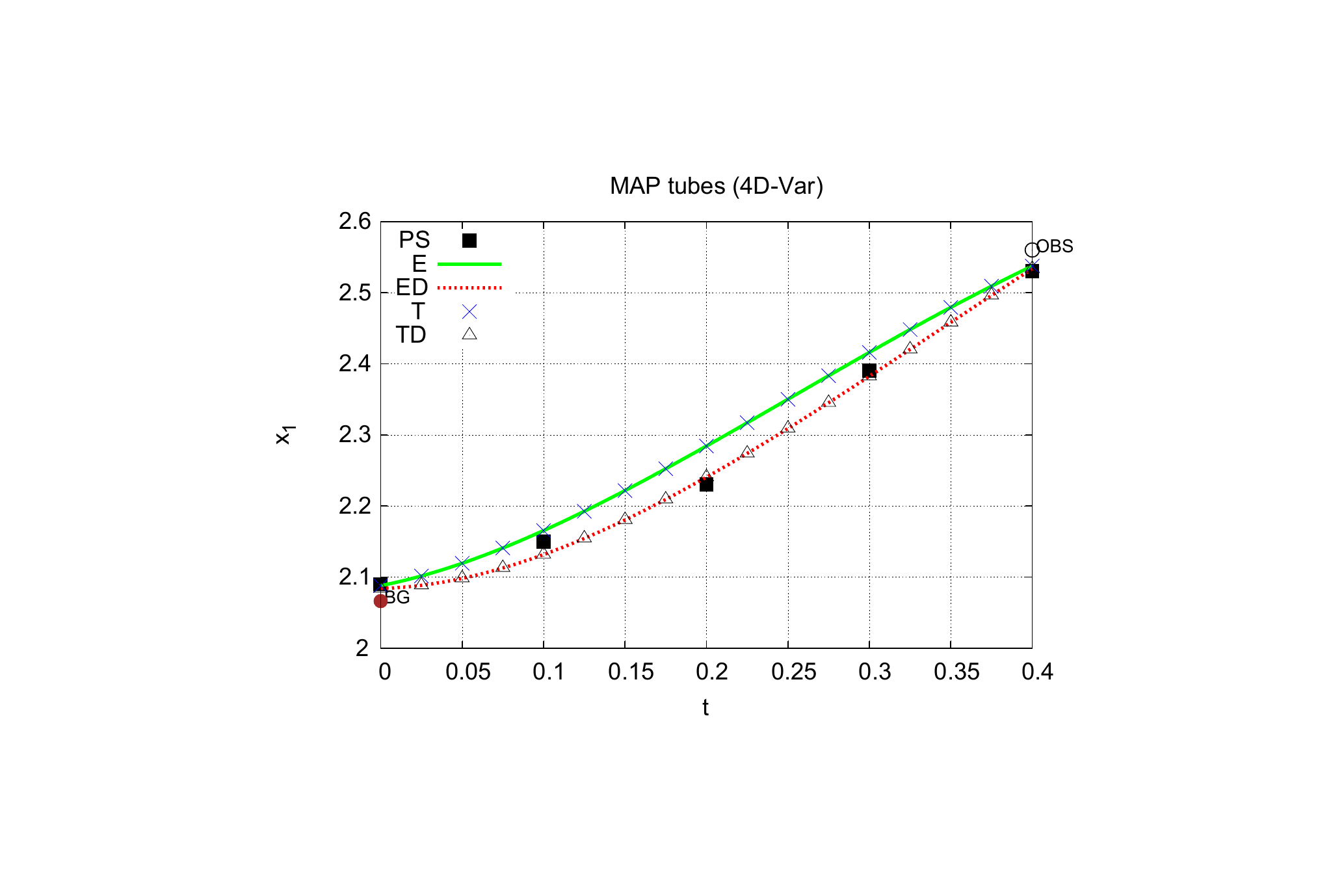}
  \caption{Most probable tube derived by 4D-Var (R\"ossler model).
    \label{MAPRoss}}
\end{figure}

The results by MCMC and 4D-Var for the R\"ossler model are shown in
Figs.~\ref{ExpRoss} and \ref{MAPRoss}, respectively. The state variable $x_1$
is chosen for the vertical axes. PS is performed with $3\times 10^{12}$
particles. The curve for PS in Fig.~\ref{MAPRoss} indicates $\hat{\phi} =
\mathrm{argmax}_{\phi}P(\phi|y)$, where $U$ represents the tube centred at
$\phi$ with radius $0.03$.

Figure~\ref{ExpRoss} shows that, just as for the hyperbolic model, E and TD
provide the proper expected path. Figure~\ref{MAPRoss} shows that ED and TD
provide the proper MAP estimate.

\begin{table*}[t]
  \caption{Applicable OM schemes. \label{schemes}}
  \begin{tabular}{llcc}
    \tophline
    & & with $\mathrm{div}\,(f)$& without $\mathrm{div}\,(f)$\\
    \middlehline
    Sampling by MCMC&   Euler scheme       &          & $\checkmark$\\
    & trapezoidal scheme & $\checkmark$ &           \\
    MAP estimate by 4D-Var&    Euler scheme & $\checkmark$ &           \\
    &  trapezoidal scheme & $\checkmark$ &      \\
    \bottomhline
  \end{tabular}
\end{table*}

\subsection{Towards application to large systems}

When one computes the cost value $J(x)$, the negative logarithm of the
posterior probability, in data assimilation, the value $f(x)$ is explicitly
computed by the numerical model while $\mathrm{div}\, f(x)$ is not. If the
dimension $D$ of the state space is large, and $f$ is complicated, the
algebraic expression of $\mathrm{div}\, f(x)$ can be difficult to obtain. The
gradient of the cost function $\nabla J(x)$ contains the derivative of
$f(x)$, which can be implemented as the adjoint model by symbolic
differentiation \citep[e.g.][]{giering1998recipes}. However, schemes with the
divergence term require the calculation of the second derivative of $f(x)$,
for which the algebraic expression can be even more difficult to obtain.
Still, there may be a way to circumvent this difficulty by utilising
Hutchinson's trace estimator \citep{hutchinson1990stochastic} (see Appendix
\ref{app_alt}). It is also clear that the Euler scheme without the divergence
term is more convenient for implementing path sampling, because it does not
require cumbersome calculation of the divergence term.

\hack{\newpage}
\conclusions

We examined several discretisation schemes of the OM functional,
$\frac1{2\sigma^2} \left( \frac{\mathrm{d}x}{\mathrm{d}t}-f(x)\right)^2 +\frac12
\mathrm{div}\,{(f)}$, for the non-linear smoothing problem
\begin{align*}
  \mathrm{d}x_t &= f(x_t) \mathrm{d}t + \sigma \mathrm{d}w_t,\\
  x_0&\sim \mathcal{N}(x_\mathrm{b},\sigma_\mathrm{b}^2 I), \quad
  (\forall m \in M)~y_m|x_m \sim \mathcal{N}(x_m,\sigma_\mathrm{o}^2 I),
\end{align*}
by matching the answers given by MCMC and 4D-Var with that given by PS,
taking the hyperbolic model and the R\"ossler model as examples.
Table~\ref{schemes} lists the discretisation schemes which were found to be
applicable, i.e. those expected to converge to the same result as the
reference solution. These results are consistent with the literature
\citep[e.g.][]{ISI:000247587700006,
  PhysRevE.94.042131,
  dutra2014maximum,stuart2004}.

This justifies, for instance, the use of the following cost function for the
MAP estimate given by 4D-Var:
\begin{align*}
  J&=
  \frac{\left|\phi_0-x_\mathrm{b}\right|^2}{2\sigma_\mathrm{b}^2}
  +
  \sum_{m\in M} \frac{\left|\phi_m-y_m\right|^2}{2\sigma_\mathrm{o}^2}
  \nonumber\\
  &+
  \delta_t \sum_{n=1}^N
  \left(
  \frac{1}{2\sigma^2}\left|
  \frac{\phi_n-\phi_{n-1}}{\delta_t}-f(\phi_{n-1})
  \right|^2
  +
  \frac{1}{2}
  \mathrm{div}\, f(\phi_{n-1})
  \right),
\end{align*}
where $n$ is the time index, $\delta_t$ is the time increment, $x_\mathrm{b}$
is the background value, $\sigma_\mathrm{b}$ is the standard deviation of the
background value, $y$ is the observational data, $\sigma_\mathrm{o}$ is the
standard deviation of the observational data, and $\sigma$ is the noise
intensity. However, the divergence term above should be excluded for the
assignment of path probability in MCMC.

For application in large systems, the Euler scheme without the divergence
term is preferred for path sampling because it does not require cumbersome
calculation of the divergence term. In 4D-Var, the divergence term can be
incorporated into the cost function by utilising Hutchinson's trace
estimator.

\codeavailability{The code for data assimilation is available at
  \url{https://github.com/nozomi-sugiura/OnsagerMachlup/}.}

\hack{\clearpage}

\appendix

\section{Scaling of the terms}\label{scale}

Taylor expansion of the $f(\psi_{n-1})$ term around $\psi_{n-\frac12}$
in scheme E gives
\begin{align*}
  \widetilde{\mathrm{OM}}&\simeq \sum_{n=1}^N \delta_t \left\{
  \sigma^{-2} \left[ \frac{\psi_n-\psi_{n-1}}{\delta_t}-f(\psi_{n-\frac12})
    -(\psi_n-\psi_{n-1})\right.\right.\nonumber\\&\left.\left.\frac{\partial f}{\partial x}(\psi_{n-\frac12})
    \right]^2
  + \mathrm{div}\,{(f)}
  \right\}\\
  &=\delta_t\left\{
  \sigma^{-2}(\text{noise}+\text{shift})^2+\text{divergence}
  \right\}.\\
  \text{noise} &\equiv \frac{\psi_n-\psi_{n-1}}{\delta_t}-f(\psi_{n-\frac12}),
 \nonumber\\& \text{shift} \equiv (\psi_n-\psi_{n-1})\frac{\partial f}{\partial x}(\psi_{n-\frac12}),
  \text{divergence} \equiv \mathrm{div}\,{(f)},
\end{align*}
where we assume order-one fluctuations, $\sigma=O(1),$ and the symbol $\psi$
represents either $\phi$ for a smooth curve or $x$ for a sample path.

For a sample path of the stochastic process, the scaling
is
$\psi_n-\psi_{n-1}=O(\delta_t^{\frac12})$, which leads to
\begin{align}
  \widetilde{\mathrm{OM}}&=\sum \delta_t\left\{\sigma^{-2}\left(
  \underbrace{\text{noise}^2}_{\delta_t^{-1}}
  +
  \underbrace{\text{noise}\times \text{shift}}_{1}
  +
  \underbrace{\text{shift}^2}_{\delta_t}
  \right)
  \right.\nonumber\\&\left.+
  \underbrace{\text{divergence}}_{1}
  \right\}.
\end{align}
The shift term induces a Jacobian that coincides with
the divergence term in TD \citep{zinn2002quantum}.

In the case of a smooth curve, there is no stochastic term, and thus
$\psi_n-\psi_{n-1}$ is the product of a bounded function $f(\psi_{n-1})$ and
$\delta_t$, which results in a value with $O(\delta_t)$. This leads to
\begin{align}
  \widetilde{\mathrm{OM}}&=\sum \delta_t\left\{\sigma^{-2}\left(
  \underbrace{\text{noise}^2}_{1}
  +
  \underbrace{\text{noise}\times \text{shift}}_{\delta_t}
  +
  \underbrace{\text{shift}^2}_{\delta_t^2}
  \right)
\right.\nonumber\\&\left. +
  \underbrace{\text{divergence}}_{1}\right\}.
\end{align}
The shift term is negligible,
but the divergence term is not.

\section{Divergence term}
\subsection{Divergence term in a trapezoidal scheme}\label{div_in_tra}

Consider two stochastic processes (cf. Sect.~6.3.2 of
\citealp{law2015data2}):
\begin{align}
  \mathrm{d}x_t &= f(x_t) \mathrm{d}t + \mathrm{d}w_t, \quad x(0) = x_0, \label{sp1}\\
  \mathrm{d}x_t &= \mathrm{d}w_t, \quad x(0) = x_0, \label{sp2}
\end{align}
where Eq.~(\ref{sp1}) has measure $\mu$ and Eq.~(\ref{sp2}) has measure
$\mu_0$ (Wiener measure). By the Girsanov theorem, the Radon--Nikodym
derivative of $\mu$ with respect to $\mu_0$ is
\begin{align}
  \frac{\mathrm{d}\mu}{\mathrm{d}\mu_0}
  &=
  \exp{\left[
      -\int_0^T \left( \frac12 |f(x)|^2 \mathrm{d}t - f(x)\cdot \mathrm{d}x \right)
      \right] }. \label{gir1}
\end{align}
If we define $F(x_T)-F(x_0)=\int_{x_0}^{x_T} f(x)\circ \mathrm{d}x$ with the
Stratonovich integral, then by Ito's formula,
\begin{align}
  \mathrm{d}F = f \cdot \mathrm{d}x + \frac12 \mathrm{div}\,{(f)} \mathrm{d}t.\label{ito1}
\end{align}
Eliminating $f\cdot \mathrm{d}x$ in Eq.~(\ref{gir1}) using Eq.~(\ref{ito1}), we obtain
\begin{align}
  \frac{\mathrm{d}\mu}{\mathrm{d}\mu_0}
  &=
  \exp\left[
      - \int_0^T \frac12 |f(x)|^2 \mathrm{d}t
      +F(x_T)-F(x_0)
    \right.\nonumber\\&\left.  -\frac12 \int_0^T \mathrm{div}\,{(f)} \mathrm{d}t
      \right] . \label{gir2}
\end{align}
Substituting $F(x_T)-F(x_0)=\int_0^T f\circ \frac{\mathrm{d}x}{\mathrm{d}t} \mathrm{d}t,$
\begin{align}
  \frac{\mathrm{d}\mu}{\mathrm{d}\mu_0}
  &=
  \exp\left[
      - \int_0^T \frac12 |f(x)|^2 \mathrm{d}t
      +\int_0^T f\circ \frac{\mathrm{d}x}{\mathrm{d}t} \mathrm{d}t
\right. \nonumber\\& \left.-\frac12 \int_0^T \mathrm{div}\,{(f)} \mathrm{d}t
      \right] . \label{gir3}
\end{align}
If we write the Wiener measure formally as $ \mu_0(\mathrm{d}x) = \exp{\left[
    -\frac12 \int_0^T \left| \frac{\mathrm{d}x}{\mathrm{d}t}\right|^2 \mathrm{d}t
    \right]}\mathrm{d}x,$
we get the following from Eq.~(\ref{gir1}),
\begin{align}
  \mu(\mathrm{d}x)
  &=\exp{\left[
      -\int_0^T \frac12
      \left|\frac{\mathrm{d}x}{\mathrm{d}t}-f(x)\right|^2
      \mathrm{d}t
      \right]}\mathrm{d}x,\label{om2}
\end{align}
and the following from Eq.~(\ref{gir3}),
\begin{align}
  \mu(\mathrm{d}x)
  &\hack{\hbox\bgroup\fontsize{8.5}{8.5}\selectfont$\displaystyle}=\exp{\left[
      -\int_0^T \frac12 \left(
      \left|\frac{\mathrm{d}x}{\mathrm{d}t}-f(x)\right|^2 +\mathrm{div}\,{(f)}
      \right) \mathrm{d}t
      \right]}\mathrm{d}x,\hack{$\egroup}\label{om1}
\end{align}
where the integrals should be interpreted in the Ito sense and in the
Stratonovich sense, respectively.

\subsection{Divergence term for smooth tube}\label{Girsa}

When weight is assigned to smooth tubes, there should always be a divergence
term, for the following reason.

Let $x$ be a diffusion process that follows the stochastic differential equation
\begin{align}
  \mathrm{d}x_t = f(x_t) \mathrm{d}t + \mathrm{d}w_t,
\end{align}
where $w$ is a Wiener process. To investigate paths near a smooth curve
$\phi$, let us consider the following stochastic process $x_t-\phi(t)$
\citep{ikeda2014stochastic,zeitouni1989onsager}:
\begin{align}
\hack{\hbox\bgroup\fontsize{8.5}{8.5}\selectfont$\displaystyle}
\mathrm{d}(x_t-\phi(t))=
  (f(x_t-\phi(t)+\phi(t))-\dot{\phi}(t) )\mathrm{d}t+\mathrm{d}w_t.\hack{$\egroup}
\end{align}
This means that if a drift $f$ is applied to the Wiener process, and the
reference frame is shifted by $\phi$, the process $x_t-\phi(t)$ which has the
drift $f(\cdot+\phi)-\dot{\phi}$ is obtained. The weight relative to the
Wiener measure can be calculated by Girsanov's formula as follows.
\begin{align}
  I_{\epsilon}(\phi)
  &\equiv
  \frac{P(\| x-\phi \|_T<\epsilon)}{P(\| w \|_T<\epsilon)}\nonumber\\
  &=
  \mathbb{E}\left[
    \exp\left(
      \int_0^T \left(f(w_t+\phi(t))-\dot{\phi}(t) \right)\right.\right.\nonumber\\&\left.\left. \cdot \mathrm{d}w_t
      -\frac12 \int_0^T \left|f(w_t+\phi(t))-\dot{\phi}(t)\right|^2\mathrm{d}t
      \right)\right.\nonumber\\&\left.
 \Big|
    \| w \|_T<\epsilon
    \right],\label{PoverP1}
\end{align}
where the expectation is taken with respect to the Wiener process $w$
conditioned to $\| w \|_T \equiv \sup_{0<t<T} |w_t|<\epsilon.$ We are going
to evaluate the terms containing $w_t$ in the exponent on the RHS of
Eq.~(\ref{PoverP1}).
\begin{enumerate}
\item \label{term_1}
  If we assume $\phi$ is a twice continuously differentiable
  function,
  then by applying Ito's product rule
  to $\dot{\phi}(t)\cdot w_t$, and using $(\forall t)~|w_t|<\epsilon$,
  \begin{align}
   \hack{\hbox\bgroup\fontsize{8.5}{8.5}\selectfont$\displaystyle} \left| \int_0^T \dot{\phi}(t)\cdot \mathrm{d}w_t \right|
    =
    \left|
    \dot{\phi}(T) \cdot w_T - \int_0^T w_t\cdot \ddot{\phi}(t) \mathrm{d}t
    \right|
    \leq A_1 \epsilon,\hack{$\egroup}
  \end{align}
where $A_1$ is a positive constant independent of $\epsilon$.
\item \label{term_2}
If we assume $f$ is a twice continuously differentiable
  function, then by using $(\forall t)~|w_t|<\epsilon$,
  \begin{align}
 \hack{\hbox\bgroup\fontsize{8.5}{8.5}\selectfont$\displaystyle}   \left|
    \int_0^T f(w_t+\phi(t))\cdot \dot{\phi}(t) \mathrm{d}t
    -
    \int_0^T f(\phi(t)) \cdot \dot{\phi}(t) \mathrm{d}t
    \right|
    \leq
    A_2 \epsilon,\hack{$\egroup}
  \end{align}
where $A_2$ is a positive constant independent of $\epsilon$.
\item \label{term_3}
In the similar manner as in 1,
  \begin{align}
   \hack{\hbox\bgroup\fontsize{8.5}{8.5}\selectfont$\displaystyle}   \left|
    \int_0^T \left|f(w_t+\phi(t)) \right|^2 \mathrm{d}t
    -
    \int_0^T \left| f(\phi(t)) \right|^2 \mathrm{d}t
    \right|
    \leq
    A_3 \epsilon,\hack{$\egroup}
  \end{align}
  where $A_3$ is a positive constant independent of $\epsilon$.
\item \label{term_4}
  The evaluation of $\int_0^T f(w_t+\phi(t)) \cdot \mathrm{d}w_t$ is as follows.
  \begin{enumerate}
  \item[a.] \label{term_4.1}
    By applying Taylor's expansion to
    $f(w_t+\phi(t))$,
    \begin{align}
     & \int_0^T f(w_t+\phi(t)) \cdot \mathrm{d}w_t
      =
      \int_0^T f(\phi(t)) \cdot \mathrm{d}w_t
      \nonumber\\&\quad\hack{\hbox\bgroup\fontsize{8.5}{8.5}\selectfont$\displaystyle} +
      \int_0^T (w_t \cdot \nabla) f(\phi(t)) \cdot \mathrm{d}w_t
      +
      \int_0^T O(w^2) \cdot \mathrm{d}w_t.\hack{$\egroup}\label{eq_4_1}
    \end{align}
  \item[b.] \label{term_4.2}
    By applying Ito's product rule
    to $w_t\cdot f(\phi(t))$, and using $(\forall t)~|w_t|<\epsilon$,
    \begin{align}
      &\int_0^T f(\phi(t)) \cdot \mathrm{d}w_t
      =
      w_T \cdot f(\phi(T))
     \nonumber\\&\quad -
      \int_0^T \sum_{i,j} w_t^i \frac{\partial f_i}{\partial
      x_j}(\phi(t)) \dot{\phi}_j(t) \mathrm{d}t=O(\epsilon).
    \end{align}
  \item[c.] \label{term_4.3}
    Regarding the second term on the RHS of
    Eq.~(\ref{eq_4_1}), we see that
    \begin{align}
      &\int_0^T (w_t \cdot \nabla) f(\phi(t)) \cdot \mathrm{d}w_t
      +\frac12 \int_0^T \nabla \cdot f(\phi(t)) \mathrm{d}t \nonumber\\
      &\hack{\hbox\bgroup\fontsize{9.5}{9.5}\selectfont$\displaystyle}=
      \int_0^T \sum_{i,j}
      \frac{\partial f_i}{\partial x_j}(\phi(t))
      w^j_t \mathrm{d}w^i_t
      +\frac12 \int_0^T \sum_{i,j} \delta_{ij}
      \frac{\partial f_i}{\partial x_j}(\phi(t)) \mathrm{d}t\hack{$\egroup} \nonumber\\
      &=
      \int_0^T \sum_{i,j}
      \frac{\partial f_i}{\partial x_j}(\phi(t))
      \left(w^j_t \mathrm{d}w^i_t+\frac12 \delta_{ij}\mathrm{d}t\right)
    \nonumber\\&\quad  =
      \int_0^T \sum_{i,j}
      \frac{\partial f_i}{\partial x_j}(\phi(t))
      \mathrm{d}\zeta^{ji}_t,
    \end{align}
    where $\zeta^{ji}_t=\int_0^t w^j_s \circ \mathrm{d}w^i_s$
    (Stratonovich integral).
  \end{enumerate}
\end{enumerate}
By applying evaluations (1)--(4) to Eq.~(\ref{PoverP1}), we obtain
\begin{align}
  I_{\epsilon}(\phi)
  &\hack{\hbox\bgroup\fontsize{8.5}{8.5}\selectfont$\displaystyle}=
  \exp{\left(-\frac12
    \int_0^T
    \left|
    f(\phi(t))-\dot{\phi}(t)
    \right|^2 \mathrm{d}t
    -\frac12 \int_0^T \nabla \cdot f(\phi(t)) \mathrm{d}t
    \right)}\hack{$\egroup}\nonumber \\
  &\hack{\hbox\bgroup\fontsize{8.5}{8.5}\selectfont$\displaystyle}\times
  \mathbb{E}\left[
    \exp\left(
      O(\epsilon)+O(\epsilon^2)
      +\int_0^T
      \sum_{i,j} \frac{\partial f_j}{\partial
      x_i}(\phi(t)) \mathrm{d}\zeta^{ji}_t
    \right.\right.\hack{$\egroup}\nonumber\\&\hack{\hbox\bgroup\fontsize{8.5}{8.5}\selectfont$\displaystyle}\left.\left.  +\int_0^T O(|w|^2) \cdot \mathrm{d}w_t
      \right)
 \Big|
    \| w \|_T<\epsilon
    \right].\hack{$\egroup}\label{I_orders}
\end{align}
On pages 450--451 in \citet{ikeda2014stochastic},
it is shown that
\begin{align}
\hack{\hbox\bgroup\fontsize{8.5}{8.5}\selectfont$\displaystyle}
\mathbb{E}\left[
    \exp{\left(
      c\int_0^T
      \sum_{i,j} \frac{\partial f_j}{\partial
      x_i}(\phi(t)) \mathrm{d}\zeta^{ji}_t
      \right)}
 \Big|
    \| w \|_T<\epsilon
    \right] \hack{$\egroup} &\
  \xrightarrow{\epsilon\to 0}
  1 \quad (\forall c),\label{conv_2}\\
  \mathbb{E}\left[
    \exp{\left(
      c\int_0^T O(|w|^2) \cdot \mathrm{d}w_t
      \right)}
 \Big|
    \| w \|_T<\epsilon
    \right] &
  \xrightarrow{\epsilon\to 0}
  1 \quad (\forall c),\label{conv_1}
\end{align}
and it is obvious that
\begin{align}
 \hack{\hbox\bgroup\fontsize{8.5}{8.5}\selectfont$\displaystyle} \mathbb{E}\left[
    \exp{\left(
      c O(\epsilon)+ c O(\epsilon^2)
      \right)}
 \Big|
    \| w \|_T<\epsilon
    \right]\hack{$\egroup} &
  \xrightarrow{\epsilon\to 0}
  1 \quad (\forall c).\label{conv_0}
\end{align}
They also showed that
if
\begin{align}
  \mathbb{E}\left[
    \exp{\left(
      c a_j
      \right)}
 \Big|
    \| w \|_T<\epsilon
    \right] &
  \xrightarrow{\epsilon\to 0}
  1\quad (\forall c) \label{each_Y}
\end{align}
for $j=1,2,\cdots,J$, then
\begin{align}
  \mathbb{E}\left[
    \exp{\left(
      \sum_{j=1}^J a_j
      \right)}
 \Big|
    \| w \|_T<\epsilon
    \right] &
  \xrightarrow{\epsilon\to 0}
  1.\label{sum_Y}
\end{align}
By applying this to Eqs.~(\ref{conv_1}), (\ref{conv_2}), and (\ref{conv_0}),
we deduce from Eq.~(\ref{I_orders}) that
\begin{align}
 & I_{\epsilon}(\phi)
  \xrightarrow{\epsilon\to 0}
 \nonumber\\&\hack{\hbox\bgroup\fontsize{8.5}{8.5}\selectfont$\displaystyle} \exp{ \left(-\frac12
    \int_0^T
    \left|
    f(\phi(t))-\dot{\phi}(t)
    \right|^2 \mathrm{d}t
    -\frac12 \int_0^T \nabla \cdot f(\phi(t)) \mathrm{d}t
    \right)}.\hack{$\egroup}\label{I_limit}
\end{align}
From evaluation (4), we also have that
\begin{align}
& \mathbb{E}\left[\exp{\left(
      \int_0^T f(w_t+\phi(t)) \cdot \mathrm{d}w_t \right) } \Big|
    \| w \|_T < \epsilon \right]\nonumber\\&\quad
  \xrightarrow{\epsilon\to 0}
  \exp{\left[-\frac12 \int_0^T \mathrm{div}\,{f}(\phi(t)) \mathrm{d}t \right]}.
  \label{divformula}
\end{align}

Equation~(\ref{divformula}) serves as an evaluation formula for the
divergence term along $\phi$ by ensemble calculation if we interpret the
expectation as an ensemble average:
\begin{align}
  \ln\mathbb{E}&\left[\exp{\left(
        \int_0^T f(w_t+\phi(t)) \cdot \mathrm{d}w_t \right) } \Big|
      \| w \|_T < \epsilon \right]
  \xrightarrow{\epsilon\to 0}
  \nonumber\\&\quad -\frac12 \int_0^T \mathrm{div}\,{f}(\phi(t)) \mathrm{d}t.
  \label{divformula2}
\end{align}
The ensemble can be generated by using a Wiener process limited to the small
area $\|w\|_T < \epsilon$. Taking the derivative of Eq.~(\ref{divformula2})
with respect to $\phi_i(t)$, we also obtain the formula for evaluating the
derivative of the divergence term along $\phi$, as follows.
\begin{align}
 & \frac{
    \mathbb{E}\left[
      \nabla f(\phi+w) \cdot \mathrm{d}w \exp{\left(
        \int_0^T f(\phi+w)\cdot \mathrm{d}w
        \right)}
 \Big|
      \|w\|_T < \epsilon
      \right]
  }
       {
         \mathbb{E}\left[
           \exp{\left(
             \int_0^T f(\phi+w)\cdot \mathrm{d}w
             \right)}
 \Big|
           \|w\|_T < \epsilon
           \right]
       }\nonumber\\&
       \xrightarrow{\epsilon\to 0}
       -\frac12 \nabla (\mathrm{div}\,{f})\mathrm{d}t, \label{graddivf}
\end{align}
where $(\nabla f(\phi+w), \mathrm{d}w)=\sum_j \frac{\partial
  f_j(\phi+w)}{\partial \phi_i} \mathrm{d}w_j$ can be calculated using
the adjoint model $\nabla f(\phi+w)$. Although these evaluation
formulas~(\ref{divformula2}) and (\ref{graddivf}) illustrate the meaning of
the divergence term, they seem too expensive to be used in the 4D-Var
iterations.

\section{Estimator for the divergence term}\label{app_alt}

Cost functions in Eqs.~(\ref{Jt_ED}) and (\ref{Jt_TD}) utilise the derivative
of the drift term $f(x)$, and thus the gradient of the term contains the
second derivative of $f(x)$, whose algebraic form is difficult to obtain in
high-dimensional systems. Here, we propose an alternative form using
Hutchinson's trace estimator \citep{hutchinson1990stochastic}, which
approximates the trace of matrix $\mathbb{E}[\xi^T A \xi]=\mathrm{tr}{(A)}$
using a stochastic vector whose components are independent, identically
distributed stochastic variables that take value $\pm 1$ with probability
$0.5$.

A realisation of the cost function is given as
\begin{align}
  \hat{J}_{\text{tube}}&(\phi|y)=
  \frac1{2\sigma_\mathrm{b}^2}\left|\phi_0-x_\mathrm{b}\right|^2
  +
  \sum_{m\in M} \frac1{2\sigma_\mathrm{o}^2}\left|\phi_m-y_m\right|^2
  \nonumber\\
  &+\delta_t \sum_{n=1}^N
  \left(
  \frac{1}{2\sigma^2}\left|
  \frac{\phi_n-\phi_{n-1}}{\delta_t}-f(\phi_{n-1})
  \right|^2\right.
 \nonumber\\&\left. +
  \frac{1}{2}
  \xi_{n-1}^T b^{-1}\left[ f(\phi_{n-1}+b\xi_{n-1})-f(\phi_{n-1})\right]
  \right),\label{Jt_ED2}
\end{align}
where $b$ is a small number. Note that $\hat{J}_{\text{tube}}(\phi|y)$ is a
stochastic variable that satisfies
\begin{align}
  \mathbb{E}\left[\hat{J}_{\text{tube}}(\phi|y)\right]&=
  J_{\text{tube}}(\phi|y).
\end{align}
If the adjoint of $f$ is at hand, the gradient of the stochastic cost
function is given as
\begin{align}
  \nabla_{\phi_n}&\hat{J}_{\text{tube}}(\phi|y)=
  \frac1{\sigma_\mathrm{b}^2}(\phi_0-x_\mathrm{b})\delta_{0,n}
  +
  \sum_{m\in M} \frac1{\sigma_\mathrm{o}^2}(\phi_m-y_m)\delta_{m,n}
  \nonumber\\
  &+
  \frac{1}{\sigma^2}
  \left(
  \frac{\phi_n-\phi_{n-1}}{\delta_t}-f(\phi_{n-1})
  \right) \qquad \qquad (n>0)\nonumber\\
  &+\frac{\delta_t}{\sigma^2}
  \left(
  -\frac1{\delta_t}-\left(\frac{\partial f}{\partial \phi_n}(\phi_n)\right)^T
  \right)
  \left(\frac{\phi_{n+1}-\phi_{n}}{\delta_t}-f(\phi_{n})\right)
 \nonumber\\& \quad (n<N) \nonumber\\
  &\hack{\hbox\bgroup\fontsize{9.5}{9.5}\selectfont$\displaystyle}+
  \frac{\delta_t}2
  \left[
    \left(\frac{\partial f}{\partial \phi_n}(\phi_n+b\xi_n)\right)^T b^{-1}\xi_n
    -
    \left(\frac{\partial f}{\partial \phi_n}(\phi_n)\right)^T b^{-1}\xi_n
    \right].\hack{$\egroup}\nonumber\\&\quad  (n<N)\label{gradJt_ED2}
\end{align}
The iterations similar to Eq.~(\ref{itr_4dvar}), $ \phi^{(k+1)} =
\phi^{(k)}-\alpha \nabla \hat{J}_{\text{tube}}, $ will work.

\hack{\clearpage}

\competinginterests{The author declares that she has no conflict of
interest.}

\begin{acknowledgements}
 The author is grateful to the referees for their comments
 which helped improve the readability of the paper.
  This work was partly supported by MEXT KAKENHI
  Grant-in-Aid for Scientific Research on Innovative Areas JP15H05819.
 All the numerical simulations were
 performed on the JAMSTEC SC supercomputer
 system.\hack{\newline}\hack{\newline}Edited by: Zoltan Toth \hack{\newline}
Reviewed by: two anonymous referees
\end{acknowledgements}

\end{document}